\documentclass[a4paper]{article}
\usepackage{epsfig}
\usepackage{graphicx}  

\date{\today}
\newcommand{\bmat}{\left(\begin{array}}
\newcommand{\emat}{\end{array}\right)}
\newcommand{\be}{\begin{equation}}
\newcommand{\ee}{\end{equation}}
\newcommand{\bea}{\begin{eqnarray}}
\newcommand{\eea}{\end{eqnarray}}

\def    \be            {\begin{equation}}
\def    \ee            {\end{equation}}
\def    \bea           {\begin{eqnarray}}
\def    \eea           {\end{eqnarray}}
\def\eps{\epsilon}
\def\a{\alpha}
\def\b{\beta}
\def\g{\gamma}
\def\d{\delta}
\def\n{\nu}

\def\lam{\lambda}
\def\th{\theta}

\def\m{\mu}

\def\d{\delta}

\def\s{\sigma}
\def\r{\rho}
\def\t{\theta}

\def\la{\lambda}

\def\eps{\epsilon}
\def\al{\alpha}
\def\b{\beta}
\def\g{\gamma}
\def\d{\delta}

\def\n{\nu}

\def\lam{\lambda}
\def\th{\theta}

\begin{document}
\title{\Large\bf The $\left(Z_2\right)^3$ symmetry of the non-tri-bimaximal pattern for the neutrino mass matrix}
\author
{ \it \bf  E. I. Lashin$^{1,2,3}$\thanks{elashin@ictp.it} , M. Abbas$^{1,2}$ \thanks{mabbas1978@yahoo.com}, N.
Chamoun$^{4,5}$\thanks{nchamoun@th.physik.uni-bonn.de}, and  S. Nasri$^{6}$ \thanks{snasri@uaeu.ac.ae} ,
\\ \small$^1$ Ain Shams University, Faculty of Science, Cairo 11566,
Egypt.\\
\small$^2$ Centre for Theoretical Physics, Zewail City of Science and Technology,\\
\small Sheikh Zayed, 6 October City, 12588, Giza, Egypt.\\
\small$^3$ The Abdus Salam ICTP, P.O. Box 586, 34100 Trieste, Italy.  \\
\small$^4$  Physics Department, HIAST, P.O.Box 31983, Damascus,
Syria. \\
 \small$^5$  Physikalisches Institut der Universit$\ddot{a}$t Bonn, Nu${\ss}$alle 12, D-53115 Bonn, Germany.
 \\
\small$^6$  Department of Physics, UAE University, P.O.Box 17551,
Al-Ain, United Arab Emirates.  }

\maketitle
\begin{center}
\small{\bf Abstract}\\[3mm]
\end{center}
In view of the recent neutrino oscillation data pointing to a non-vanishing value for the
 smallest mixing angle ($\theta_z$), we derive and find explicit realizations of the $\left(Z_2\right)^3$ flavor symmetry which characterizes,
 for the neutrino mass matrix, uniquely a variant of the tripartite form, originally conceived to lead to the tri-bimaximal mixing
 with $\theta_z=0$, so that to allow now for a non-tri-bimaximal pattern with non-zero $\theta_z$.
 We impose
this flavor symmetry in a setting including the charged leptons and we see that it  can make room, through higher order terms
 involving new SM-singlet scalars, for the mass hierarchy of charged leptons. Moreover, within type-I seesaw mechanism augmented with
 the flavor symmetry, certain patterns occurring
 in both the Dirac and the Majorana neutrino mass matrices can accommodate all types of  mass hierarchies in the effective
 neutrino mass matrix, but no lepton/baryon asymmetry can be generated. Finally, we discuss how type-II seesaw mechanism, when supplemented
 with the flavor symmetry, could be used to interpret the observed baryon asymmetry through leptogenesis.
\begin{minipage}[h]{14.0cm}
\end{minipage}
\vskip 0.3cm \hrule \vskip 0.5cm

\section{Introduction}
Neutrino oscillation experiments have established that neutrino flavor and mass states do mix \cite{nakamura10}.
The three flavor states ($\n_e,\n_\m,\n_\tau$) are quantum linear combinations of the three light mass states ($\n_1,\n_2,\n_3$)
through a unitary mixing matrix $U^\nu_L$ parameterized by three rotation angles ($\t_x,\t_y,\t_z$) and three phases
$\left[\d \mbox{(Dirac phase)},\r \mbox{ and } \s \mbox{(Majorana phases)}\right]$ as follows:
\bea
{M}_\nu &=& U^\nu_L\; M^{\mbox{diag}}_\n\; (U^\nu_L)^T,\nonumber \\
M^{\mbox{diag}}_\n &=&  \mbox{diag}(m_1,m_2,m_3),\nonumber\\
U^\nu_L &=&  \left ( \matrix{ c_x c_z & s_x c_z & s_z \cr - c_x s_y
s_z - s_x c_y e^{-i\delta} & - s_x s_y s_z + c_x c_y e^{-i\delta}
& s_y c_z \cr - c_x c_y s_z + s_x s_y e^{-i\delta} & - s_x c_y s_z
- c_x s_y e^{-i\delta} & c_y c_z \cr } \right ) \mbox{diag}(e^{i\rho},e^{i\sigma},1).
\label{mixmat}
\eea
where ${M}_\nu$ is the effective neutrino mass matrix which is symmetric assuming the neutrinos are of
Majorana type, $m_i$'s are the masses of the neutrino mass states, and with $s_x \equiv \sin\theta_x$, $c_y \equiv \cos\theta_y$ and so
on.

Our convention for parameterizing the mixing matrix in Eq.~(\ref{mixmat}) is related to the standard convention used, say, in the recent data analysis of \cite{fogli}, in that the mixing angles coincide in the two conventions but with different nomenclatures; strictly speaking,
\begin{equation}
\theta_x \; \equiv \; \theta_{12} \; , ~~~~~ \theta_y \;
\equiv \; \theta_{23} \; , ~~~~~ \theta_z \; \equiv \;
\theta_{13}.
\end{equation}
As to  the phases, one needs to decompose the unitary mixing matrix $U_L^\nu$ as
\bea \label{Udecomposition} U_L^\nu &=& R_{23}(\theta_y) \cdot R_z(\delta) \cdot R_{12}(\theta_x) \cdot P\eea
where $R_{12}$ and $R_{23}$ represent rotations around the $z$ and $x$ axes respectively, whereas $R_z(\delta)$ and $P$ depend on the
convention adopted. The standard parametrization, denoted by a superscript tilde, consists of taking
\bea \label{standard convention}
\widetilde{R_z}(\delta) =
\left ( \matrix{ c_z & 0  & s_z \,e^{-i\,\d}\cr
0 & 1 & 0 \cr
- s_z\,e^{i\,\d} & 0 &  c_z\cr } \right )
&,&
 \widetilde{P} = \mbox{diag}(1,e^{i{\phi_2\over 2}},e^{i{(\phi_3+2\d)\over2}}).
\eea 
where $\phi_2$ and $\phi_3$ designate the Majorana phases. On the other hand, in the parametrization we adopt, which has the advantage
that the Dirac phase $\d$ deos not appear in the effective mass term of the neutrinoless double beta decay \cite{Xing}, we have
\bea
R_z(\delta) =
\left ( \matrix{ c_z & 0  & s_z \cr
0 & e^{-i\,\d} & 0 \cr
-s_z & 0 &  c_z\cr } \right ),
&,&
P = \mbox{diag}(e^{i\rho},e^{i\sigma},1).
\label{our parametrization}
\eea
One can show, by calculating some rephasing invariant quantities \cite{oztex}, that  
the Dirac phase is the same for both parameterizations while the Majorana phases in the two parameterizations are related through simple linear relations: \bea
\rho = -{\phi_3\over 2}, & & \sigma = {\phi_2 -\phi_3\over 2}.
\eea

Likewise, one can diagonalize the charged lepton mass matrix linking the left handed (LH) leptons to their right handed (RH) counterparts,
and thus being not necessarily symmetric, albeit now by a bi-unitary transformation:
\begin{equation}
{M}_l = U^l_L\; \mbox{diag}(m_e,m_\m,m_\tau)\; (U^l_R)^\dagger.
\end{equation}
 The
observed neutrino mixing matrix comes from the mismatch between
$U^l_L$ and $U^\nu_L$ in that one can put experimental constraints on the elements of the flavor mixing matrix $V=(U^l_L)^\dagger\, U^\nu_L$.
If one works in the flavor basis where $M_l M_l^\dagger$
is diagonal, then
$U_L^l = {\bf 1}$ is the identity matrix, and the whole measured flavor mixing matrix originates from  $U^\nu_L$.

The authors of \cite{HPS} noticed that, dropping the phases on which no definite experimental measurements exist up till now,
the experimental constraints impose a specific pattern named tri-bimaximal (TB):
\begin{eqnarray}
U^{\nu}_L \simeq V^{TB} &\equiv&
\pmatrix{\sqrt{2/3} & 1/\sqrt{3} & 0 \cr -1/\sqrt{6} & 1/\sqrt{3}
& 1/\sqrt{2} \cr 1/\sqrt{6} & -1/\sqrt{3} & +1/\sqrt{2}},
\end{eqnarray}
amounting to $\t_x=\arcsin(\frac{1}{\sqrt{3}}),\; \t_y=\frac{\pi}{4}$ and $\t_z=0$.

The author of \cite{matripar} showed that the TB pattern is equivalent to a certain form for the $M_\n$ in the flavor basis called `tripartite':
\bea
\label{vtb-trip}
M^{\mbox{diag}}_\n =(V^{TB})^T \; M^{TB}_\n \;V^{TB} &\Leftrightarrow&  {M}^{TB}_{\nu} = { M}_A + { M}_B + { M}_C,
\eea
where
\begin{eqnarray} \label{trip}
{M}_A = A \pmatrix {1 & 0 & 0 \cr 0 & 1 & 0 \cr 0 & 0 & 1}, ~~ {
M}_B = B \pmatrix {-1 & 0 & 0 \cr 0 & 0 & 1 \cr 0 & 1 & 0}, ~~ {
M}_C = C \pmatrix {1 & 1 & -1 \cr 1 & 1 & -1 \cr -1 & -1 & 1},
\label{formz}
\end{eqnarray}
with neutrino eigen masses related to the tripartite coefficients via:
\begin{eqnarray}
\label{neigenm}
&m_1=A-B, \; m_2 =
A-B + 3C, \; m_3 = A+B, & \nonumber\\
&A=(m_1+m_3)/2, \; B=(m_3-m_1)/2, \; C=(m_2-m_1)/3. \;&
\end{eqnarray}
We showed in \cite{z2,u1} that $M_A+M_B$ has a $U(1)$ underlying
symmetry which is broken by $M_C$, however a residual $\left(Z_2\right)^3$ symmetry
is left unbroken
and this remaining symmetry (called henceforth S-symmetry) characterizes uniquely the tripartite
form, and hence the TB pattern. More explicitly, we determined the set of invertible matrices $S$ which leave the neutrino
mass matrix corresponding to the TB pattern unchanged under congruency (or what was called in \cite{maform} the form invariance condition):  
\begin{equation}
\label{form}
S^T \;{M_\n}\;S = {M_\n}
\end{equation}
We have found that this set includes eight elements forming a subgroup of the nonsingular $3\times 3$ complex matrices under
matrix multiplication. Moreover, this subgroup is isomorphic to the multiplicative group $(Z_2)^3$ with three generators, whence the name. 

From a mathematical viewpoint, the form of $V^{TB}$ is just a unitary matrix whose
elements are expressed in terms of simple numbers via algebraic operations. One would wonder thus what would be the consequences
on the flavor S-symmetry had the experimental data led to different values. In \cite{shaofeng}, relations between residual $Z_2$
symmetries and patterns not conforming to the TB mixing pattern were studied. 

Actually, and despite the phenomenological success of the TB pattern \cite{TB}, it is important, in light of the recent neutrino data best fit
\cite{fogli,Double,DAYA,RENO,Valle}
amounting to $\t_x=33.58^o,\t_y=40.40^o$ and $\t_z=8.33^o$, to restudy the question of the $\left(Z_2\right)^3$ S-symmetry, possessed by the neutrino mass matrix, in the case
of TB mixing but with non-zero $\t_z$ (which we would call a non-tri-bimaximal mixing pattern and denote it by `NTB'), and find an explicit
form of this flavor S-symmetry different from the specific form we found in \cite{z2} corresponding to the TB case. The object of this
paper is to present such a study with the cosmological consequences for the baryogenesis.

For this, we follow the method detailed in \cite{z2} which amounts to finding the generators of the $\left(Z_2\right)^3$ S-symmetry in
the diagonalized basis by imposing the form invariance condition, then deducing the corresponding generators of the $\left(Z_2\right)^3$ S-symmetry in the flavor basis. We find also a `variant' tripartite form characterizing the NTB pattern for generic
$\t_z$. Although all the formulae are derived with a general $\t_z \neq 0$, we nonetheless, for numerical applications and definite discussions, would specify the NTB pattern
to the case $\t_z = \arcsin(\frac{1}{\sqrt{50}})$ matching, to a good approximation, the recent best fitted data, and would denote this 
special case by a subscript zero mark.

All the same, we stress that the special value $\t_z = \arcsin(\frac{1}{\sqrt{50}})$ is not a result from any theoretical consideration, but 
suggested only by experiment, whereas the general results are independent of the explicit value of $\t_z$. In fact, we could as well
leave the values of $\t_x$ and $\t_y$ as free parameters. However, the resulting expressions are cumbersome enough not to be of practical use,
and we have opted to keep these two mixing angles fixed to their TB values consistent with experiment. This was motivated in particular by the fact that, compared
to the previous oscillation data in \cite{fogli-old}, the new data ({\it e.g.} \cite{fogli}) did not signal a palpable
change in the values of $\t_x$ and $\t_y$ in contrast to $\t_z$ where the former data were consistent with the value $\t_z =0$ rejected now
by the present data (see Table \ref{data}).

\begin{table}[h]
 \begin{center}
{\small
 \begin{tabular}{c | c c c c | c c c}
\hline
\hline
\mbox{Parameter} &\mbox{Present Best fit} &  \mbox{Present} $2\s$ \mbox{range} & \mbox{NTB}& $\mbox{NTB}_0$&  \mbox{Past Best fit} &  \mbox{Past} $2\s$ \mbox{ range} & \mbox{TB}  \\
\hline
 $\th_x$ & $33.58^o$ &$\left[31.95^o, 36.09^o\right]$ & $35.27^o$ & $35.27^o$ & $34.08^o$ & $\left[31.11^o, 37.50^o\right]$ & $35.27^o$ \\
 $\th_y$ & $40.40^o$ &$\left[36.87^o, 50.77^o\right]$ & $45^o$ & $45^o$ & $41.55^o$ & $\left[35.86^o, 51.97^o\right]$ & $45^o$ \\
 $\th_z$ & $8.33^o$ &$\left[6.29^o, 11.68^o\right]$ & $\t_z$ & $8.13^o$ & $5.44^o$ & $\left[0^o, 10.30^o\right]$ & $0^o$ \\
\hline
 \end{tabular}
 }
 \end{center}
  \caption{\small \label{data} The present (\cite{fogli}) and past (\cite{fogli-old}) global-fit results for the three neutrino mixing angles $(\th_x, \th_y, \th_z)$, with their $2\s$ ranges ($95\%$ C.L.). We show also the corresponding values in the generic and the special NTB patterns, and in the TB pattern.}
 \end{table}

As in \cite{z2}, we show that imposing the flavor S-symmetry in a complete lepton setting, coupled to scalar fields singlet with respect to the standard model (SM) group but with suitably chosen Yukawa couplings, can lead in a natural way to the charged lepton mass hierarchy. Moreover,
by assuming the canonical type-I seesaw mechanism formula:
\begin{equation}
\label{Majmass}
{M}_\nu = { M}^D_\nu { M}_R^{-1} ({ M}^D_\nu)^T,
\end{equation}
where $M_R$ is the heavy Majorana mass matrix
for the RH neutrinos, and $M^D_\n$ is the Dirac neutrino mass matrix, one can accommodate the different possible neutrino mass
patterns provided the neutrino components transform in a specific way under the S-symmetry. However, and as was the case of the TB pattern in
\cite{z2}, we show that we need to call for type-II seesaw mechanism in order to account for the observed baryon to photon density.

The plan of the paper is as follows. In section 2 we find an explicit realization of the $\left(Z_2\right)^3$ S-symmetry leading to the
observed NTB pattern. In section 3 we impose the symmetry on the charged leptons and discuss their mass hierarchy. In section 4 we show
within type-I seesaw scenario how we can obtain the different types of neutrino mass hierarchies. We show also why we shall need
in section 5 the other type-II seesaw mechanism in order to explain the baryon asymmetry generated by lepton asymmetry. We end up
by a summary in section 6.

\section{{\large \bf The underlying symmetry of the NTB pattern}}
We review here the approach of form invariance, which we used in \cite{z2}, in order to find an explicit form of the $\left(Z_2\right)^3$ S-symmetry
leading to the NTB pattern defined by:
\be
\label{NTB pattern}
\t_y= \arcsin\frac{1}{\sqrt 2},\; \t_x= \arcsin\frac{1}{\sqrt 3},\; \t_z \neq 0,
\ee
which would correspond to a flavor mixing matrix:
\bea
V^{NTB} &=& \pmatrix{\sqrt{\frac{2}{3}} c_z & \frac{c_z}{\sqrt{3}} & s_z \cr -\frac{1}{\sqrt{6}}(1+\sqrt{2} s_z) & \frac{1}{\sqrt{3}} (1-\frac{s_z}{\sqrt{2}} )
& \frac{c_z}{\sqrt{2}} \cr \frac{1}{\sqrt{6}} (1-\sqrt{2} s_z) & -\frac{1}{\sqrt{3}} (1+\frac{s_z}{\sqrt{2}} ) & \frac{c_z}{\sqrt{2}}},
\label{VNTB}
\eea

We see that seeking unitary matrices $S$ satisfying the form invariance formula (Eq. \ref{form}) is equivalent to finding unitary matrices $U$ satisfying this invariance in the diagonalized basis:
\begin{equation}
\label{formdiag}
U^T\; M^{{\mbox{diag}}}_\n\; U = M^{{\mbox{diag}}}_\n,
\end{equation}
provided
\bea
S&=&V^*\cdot U \cdot V^T.
\eea
The different values of the $m_i$'s lead, via Eq. (\ref{formdiag}), automatically to
\bea
U=\mbox{diag}(\pm 1, \pm 1, \pm 1),
\label{U}
\eea
and thus, denoting by $I_i$
the reflection across the $i^{\mbox{th}}$ axis ($x=1,y=2,z=3$: {\it {e.g.}} $I_1\equiv I_x=\mbox{diag}(-1,+1,+1)$),
we find the symmetries $S_i$'s which characterize the NTB pattern for generic $\t_z$, 
\bea
S_1=(V^{NTB})^* \cdot \; \mbox{diag}(-1, 1, 1) \cdot \; (V^{NTB})^T &=&  \pmatrix{1-4c_z^2/3 & 2c_z(1+\sqrt{2}s_z)/3  & 2c_z(-1+\sqrt{2}s_z)/3 \cr 2c_z(1+\sqrt{2}s_z)/3 & 2(-\sqrt{2}s_z+c_z^2)/3
& c_{2z}/3 \cr  2c_z(-1+\sqrt{2}s_z)/3 & c_{2z}/3 & 2(\sqrt{2}s_z+c_z^2)/3},\nonumber\\
S_2=(V^{NTB})^* \cdot \; \mbox{diag}(1, - 1, 1) \cdot \; (V^{NTB})^T &=& \pmatrix{1-2c_z^2/3 & \sqrt{2}c_z(-\sqrt{2}+s_z)/3  & \sqrt{2}c_z (\sqrt{2}+s_z)/3 \cr \sqrt{2}c_z(-\sqrt{2}+s_z)/3 & (2\sqrt{2}s_z+c_z^2)/3
& (1+c_z^2)/3 \cr \sqrt{2}c_z (\sqrt{2}+s_z)/3 & (1+c_z^2)/3 & (-2\sqrt{2}s_z+c_z^2)/3},\nonumber \\
S_3=(V^{NTB})^* \cdot \; \mbox{diag}(1, 1, - 1) \cdot \; (V^{NTB})^T &=& \pmatrix{c_{2z} & -s_{2z}/\sqrt{2}  & -s_{2z}/\sqrt{2} \cr -s_{2z}/\sqrt{2} & s_z^2 & -c_z^2 \cr  -s_{2z}/\sqrt{2} & -c_z^2 & s_z^2}.
\eea
The three matrices ($S_i$, $i=1,2,3$) represent the generators of the $(Z_2)^3$-symmetry satisfied by the neutrino mass matrix. 

Taking into consideration that the form invariance formula involves the matrix $S$ quadratically, and that, say, $S_1 S_2 = -S_3$, we have thus proved,
for each generic value $\t_z$, the following:
  \bea
  \label{equivalence}
  \left( \forall S \in \langle S_1, S_2, S_3 \rangle, S^T \cdot M_\n \cdot S=M_\n \right) &\Leftrightarrow& \left( \forall S \in \langle S_1 , S_2 \rangle, S^T \cdot M_\n
  \cdot S = M_\n \right) \nonumber \\ &\Leftrightarrow&  \left[ (V^{NTB})^T \cdot M_\n \cdot V^{NTB} = M_\n^{\mbox{diag}}
\right].
\eea

We would like now to find a variant of the tripartite form which would characterize the generic NTB pattern so that to generalize the
equivalence formula of Eqs. (\ref{vtb-trip}, \ref{trip}). For this we note that the matrix $R_z(\d)$ (Eq. \ref{our parametrization}), when the phases are dropped, becomes $R_{13}(\theta_z)$ the rotation around the y-axis, so we have
\bea
V^{NTB} = R_{23}(\theta_y) \cdot R_{13}(\theta_z) \cdot R_{12}(\theta_x) &,& V^{TB} = V^{NTB}(\t_z=0),
\eea
where the rotation matrices are given by
\bea
R_{12}(\theta_x) \equiv \left ( \matrix{ c_x & s_x   & 0 \cr -s_x & c_x
& 0 \cr 0   & 0 & 1 \cr} \right ),  & R_{23}(\theta_y) \equiv \left ( \matrix{ 1 & 0   & 0 \cr 0 & c_y
& s_y \cr 0   & -s_y & c_y \cr} \right ) & R_{13}(\theta_z) \equiv \left ( \matrix{ c_z & 0   & s_z \cr 0 & 1
& 0 \cr -s_z   & 0 & c_z \cr} \right ),
\label{xyzr}
\eea

We get then
\bea
\label{transition}
V^{NTB} &=& W \cdot V^{TB},\nonumber\\
W &=&  R_{23}\left(\theta_y=\frac{\pi}{4}\right)\; R_{13}(\theta_z)\; R^{-1}_{23}\left(\theta_y=\frac{\pi}{4}\right)
=\left ( \matrix{ c_z      & \frac{s_z}{\sqrt{2}} & \frac{s_z}{\sqrt{2}} \cr
-\frac{s_z}{\sqrt{2}} & \frac{1+c_z}{2}   &
 -\frac{1-c_z}{2} \cr -\frac{s_z}{\sqrt{2}} &
-\frac{1-c_z}{2}  & \frac{1+c_z}{2} \cr} \right ).
\eea
Using the equality,
\be
(V^{TB})^T \cdot M^{TB}_\n \cdot V^{TB}  = (V^{NTB})^T \cdot M^{NTB}_\n \cdot V^{NTB} = M_\n^{\mbox{diag}}\equiv \mbox{diag}(m_1,m_2,m_3),
\ee
we can relate the neutrino mass matrices in the NTB and the TB patterns with each other in the following way:
\bea
\label{ntb-tb}
M^{NTB}_\n &=& W \cdot M^{TB}_\n \cdot W^T.
\eea

At this stage, it would be useful if the sought form for $M^{NTB}_\n$ characterizing uniquely the generic NTB pattern varies slightly
from the tripartite form (Eqs. \ref{vtb-trip}, \ref{trip}), and hence called hereafter a `generic' modified tripartite form. To fix the ideas, we choose a pattern where the diagonal elements of $M_\n^{NTB}$ `resemble' their counterparts in the tripartite form of $M_\n^{TB}$, in that we impose:
\bea
\label{vtrip-diag}
M_\n^{NTB}(1,1) &=& A^z-B^z+C^z,\nonumber \\
M_\n^{NTB}(2,2) &=& A^z+C^z, \nonumber\\
M_\n^{NTB}(3,3) &=& A^z-C^z.
\eea
Now, by equating the diagonal elements in both sides of Eq. (\ref{ntb-tb}), one can express the ($A^z,B^z,C^z$) in terms of the tripartite form coefficients ($A,B,C$), then by inverting these expressions one can find the non-diagonal elements of $M_\n^{NTB}$ in terms of ($A^z,B^z,C^z$):
\bea \label{vtrip-nondiag}
M_\n^{NTB}(1,2) &=& \frac{\sqrt{2}\, s_z\, c_z}{1-3\,s_z^2}B^z+\frac{c_z\, (1-2\,\sqrt{2}\,s_z)}{2\,(1-3\,s_z^2)}C^z-\frac{c_z}{\sqrt{2}\,s_z} C^z, \nonumber\\
M_\n^{NTB}(1,3) &=& \frac{\sqrt{2}\, s_z\, c_z}{1-3\,s_z^2}B^z+\frac{c_z\, (1-2\,\sqrt{2}\,s_z)}{2\,(1-3\,s_z^2)}C^z+\frac{c_z}{\sqrt{2}\,s_z} C^z,\nonumber \\
M_\n^{NTB}(2,3) &=& \frac{c^2_z}{1-3\,s_z^2}B^z+\frac{-2\,s_z\,c_z^2+2\,\sqrt{2}\,c_z^2-\sqrt{2}}{2\,s_z(1-3\,s_z^2)}C^z.
\eea
The relations in Eqs. (\ref{vtrip-diag}, \ref{vtrip-nondiag}) define the generic modified tripartite form for the symmetric neutrino mass matrix.

This pattern of $M_\n^{NTB}$ is characterized also by some `mass relations' relating its different entries:
{\small
\bea \label{ntbid}
M_\n^{NTB}(1,2) + M_\n^{NTB}(1,3) &=& {\tan{2\,\t_z}\over 2} \,\left[M_\n^{NTB}(2,3) - M_\n^{NTB}(1,1) + \frac{M_\n^{NTB}(2,2) + M_\n^{NTB}(3,3)}{2}\right],\nonumber\\
M_\n^{NTB}(2,2) - M_\n^{NTB}(3,3) &=& \sqrt{2}\,\tan{\t_z}\,\left[M_\n^{NTB}(1,3) - M_\n^{NTB}(1,2)\right],\nonumber \\
2\,M_\n^{NTB}(1,1) +6\, M_\n^{NTB}(2,2) &=& \left[M_\n^{NTB}(3,3) + M_\n^{NTB}(2,2)\right]\,\left(1+\cos{2\,\t_z}\right)
- 2\,\left[M_\n^{NTB}(1,1)-M_\n^{NTB}(2,3)\right]\, \cos{2\,\t_z}\nonumber\\
&& - 2\,\left[M_\n^{NTB}(1,2)+M_\n^{NTB}(1,3)\right]\, \cos{2\,\t_z}- 2\,\left[M_\n^{NTB}(1,2)-M_\n^{NTB}(1,3)\right]\, \cos{\t_z}\nonumber\\
&& - 2\,\left[M_\n^{NTB}(3,3)-M_\n^{NTB}(2,2)\right]\, \cos{\t_z}+ 2\,\sqrt{2}\left[M_\n^{NTB}(1,2)+M_\n^{NTB}(1,3)\right]\, \sin{2\,\t_z}\nonumber\\
&& +{1\over \sqrt{2}}\left[M_\n^{NTB}(2,2)+M_\n^{NTB}(3,3)\right]\, \sin{2\,\t_z}- \sqrt{2}\left[M_\n^{NTB}(1,1)- M_\n^{NTB}(2,3)\right]\, \sin{2\,\t_z}\nonumber\\
&& -2\,\sqrt{2}\left[M_\n^{NTB}(1,3)-M_\n^{NTB}(1,2)\right]\, \sin{\t_z}- \sqrt{2}\left[M_\n^{NTB}(3,3)- M_\n^{NTB}(2,2)\right]\, \sin{\t_z}.\nonumber\\
\label{idntbm}
\eea
}
These relations are to be contrasted with the simple ones characterizing the TB pattern ($\t_z=0$):
\bea
M_\n^{TB}(1,2) + M_\n^{TB}(1,3) &=& 0,\nonumber \\
M_\n^{TB}(2,2) - M_\n^{TB}(3,3) &=& 0,\nonumber \\
M_\n^{TB}(1,1) + M_\n^{TB}(1,2)- M_\n^{TB}(2,2) + M_\n^{TB}(2,3) &=& 0.
\label{idtbm}
\eea

Up till now, the obtained results hold independently of an explicit value of $\t_z$. However, it is useful now, in order
to contrast our analysis with the experimental data, to specify our results to the precise value ($\t_z\simeq 8.13^o\simeq\arcsin\frac{1}{\sqrt{50}}$).
In this specific case, we have the `observed' mixing matrix:
\bea
V^{NTB}_0 &=& \pmatrix{\frac{7}{15} \sqrt{3} & \frac{7}{30} \sqrt{6} & \frac{1}{10} \sqrt{2} \cr -\frac{1}{5} \sqrt{6} & \frac{3}{10} \sqrt{3}
& \frac{7}{10} \cr \frac{2}{15} \sqrt{6} & -\frac{11}{30} \sqrt{3} & \frac{7}{10}}
\label{VNTB0}
\eea
As to the $(Z_2)^3$ generators for the S-symmetry, they are given now as follows.
\bea \label{symm-real-0}
  S_{01} = \pmatrix{-\frac{23}{75} & \frac{14\sqrt{2}}{25}  & -\frac{28\sqrt{2}}{75} \cr \frac{14\sqrt{2}}{25} & \frac{13}{25}
& \frac{8}{25} \cr -\frac{28\sqrt{2}}{75} & \frac{8}{25} & \frac{59}{75}},
  S_{02} = \pmatrix{\frac{26}{75} & -\frac{21\sqrt{2}}{50}  & \frac{77\sqrt{2}}{150}  \cr -\frac{21\sqrt{2}}{50} & \frac{23}{50}
& \frac{33}{50} \cr \frac{77\sqrt{2}}{150} & \frac{33}{50} & \frac{29}{150}},
 S_{03} = \pmatrix{\frac{24}{25} & -\frac{7\sqrt{2}}{50}  & -\frac{7\sqrt{2}}{50}  \cr -\frac{7\sqrt{2}}{50}  & \frac{1}{50}
& -\frac{49}{50} \cr -\frac{7\sqrt{2}}{50}  & -\frac{49}{50} & \frac{1}{50}}.
\eea
We also find the transition matrix between the TB and our special NTB patterns as equal to 
 \bea
 \label{W0}
 W_0 &=&  \pmatrix{\frac{7\sqrt{2}}{10} & \frac{1}{10}  & \frac{1}{10}  \cr -\frac{1}{10} & \frac{7\sqrt{2}+10}{20}
&\frac{7\sqrt{2}-10}{20} \cr -\frac{1}{10} & \frac{7\sqrt{2}-10}{20} & \frac{7\sqrt{2}+10}{20}},
\eea
which corresponds to the NTB `special' modified tripartite form
\bea
\label{ctrip0}
M^{NTB}_{\n0} &=& \pmatrix{A_0-B_0+C_0 & \frac{7\sqrt{2}}{47}(B_0-22C_0) & \frac{7\sqrt{2}}{47}(B_0+25C_0)  \cr \frac{7\sqrt{2}}{47}(B_0-22C_0) & A_0+C_0
&  \frac{1}{47}(49B_0+191C_0)\cr  \frac{7\sqrt{2}}{47}(B_0+25C_0) &  \frac{1}{47}(49B_0+191C_0) & A_0-C_0},
\eea
where the neutrino eigen masses and the coefficients of this special modified tripartite form are related by:
\bea
\label{eigen-trip-0}
m_1 = A_0-\frac{49}{47}B_0+\frac{279}{47}C_0 &,& A_0 = \frac{49}{100} m_3 + \frac{101}{300} m_2 + \frac{13}{75} m_1
\nonumber, \\
m_2 = A_0-\frac{49}{47}B_0-\frac{426}{47}C_0 &,& B_0 = \frac{47}{100} m_3 - \frac{17}{300} m_2 - \frac{31}{75} m_1
\nonumber, \\m_3 = A_0+\frac{51}{47}B_0+\frac{194}{47}C_0 &,& C_0 = -\frac{1}{15} m_2 + \frac{1}{15} m_1.
\eea
The resulting mass spectra can accommodate all types of neutrino mass hierarchies as follows;
\begin{itemize}
\item{ Normal hierarchy}: with
\begin{eqnarray}
 A_0 \simeq B_0,\;\; C_0 \ll B_0,
\end{eqnarray}
we get a mass spectrum corresponding to normal hierarchy:
\begin{eqnarray} \label{nh}
[m_1,m_2,m_3] &\simeq& \left[ A_0-\frac{49}{47}B_0,\; A_0-\frac{49}{47}B_0,\;A_0+\frac{51}{47}B_0 \right].
\end{eqnarray}
\item{ Inverted hierarchy}: with
\begin{eqnarray}
A_0 \simeq - B_0,\;\; C_0 \ll B_0,
\end{eqnarray}
we get a mass spectrum corresponding to inverted hierarchy:
\begin{eqnarray} \label{ih}
[m_1,m_2,m_3] &\simeq&  \left[ A_0-\frac{49}{47}B_0,\; A_0-\frac{49}{47}B_0,\; \frac{194}{47}C_0 \right].
\end{eqnarray}
\item{ Degenerate case}: with
\begin{eqnarray}
A_0 \gg B_0 \gg C_0,\;
\end{eqnarray}
we get a quasi degenerate mass spectrum
\begin{eqnarray} \label{dh}
[m_1,m_2,m_3] &\simeq& \left[ A_0,\; A_0,\; A_0 \right].
\end{eqnarray}
\end{itemize}

In addition to what precedes, we could check that the symmetry generated by the generic ($S_1, S_2, S_3$) (or by any two of the three elements) is equivalent to the generic modified tripartite form. In particular, and limiting our attention to our special case ($\t_z=\arcsin \frac{1}{\sqrt{50}}$) we have the following equivalences corresponding to the special modified tripartite form:
\bea
\label{equivalence}
\left[ (S_{0i})^T \cdot M_{\n 0} \cdot S_{0i}=M_{\n 0} \right] & \Leftrightarrow &
\left\{ k\neq l \Rightarrow \left[ (V_0^{NTB})^T \cdot M_{\n 0} \cdot V_0^{NTB} \right]_{kl} =0 \right\}
\nonumber\\
& \Leftrightarrow & {\small \left[ \exists A_0,B_0,C_0: M_{\n0} = \pmatrix{A_0-B_0+C_0 & \frac{7\sqrt{2}}{47}(B_0-22C_0) & \frac{7\sqrt{2}}{47}(B_0+25C_0)  \cr \frac{7\sqrt{2}}{47}(B_0-22C_0) & A_0+C_0
&  \frac{1}{47}(49B_0+191C_0)\cr  \frac{7\sqrt{2}}{47}(B_0+25C_0) &  \frac{1}{47}(49B_0+191C_0) & A_0-C_0}
\right]},\nonumber\\
\eea
where $i$ spans the set $\{1,2,3\}$ or any two different elements in it.

\section{ The charged-lepton mass matrix}
Up to this point, we have found explicit realizations of the $\left(Z_2\right)^3$ S-symmetry which would lead to the
phenomenologically interesting form for $M_\n$ corresponding to the NTB pattern. However, we would now lift our findings to
an underlying symmetry level and construct a model for the leptons where the Lagrangian is kept invariant under the flavor S-symmetry.

We see that the SM term
 \bea
 \label{L1}
 {\cal{L}}_1 &=& Y_{ij} \overline{L}_i \Phi  l^c_j \,
 \eea
has to be absent provided the SM-singlet charged RH leptons $l^c_j$ and the SM Higgs are singlet under $S=\left(Z_2\right)^3$, whereas
the left doublets transform componentwise as:
\begin{equation}
 L_i \rightarrow S_{ij}L_j \;
\end{equation}
To show this, it is sufficient to note that the invariance of ${\cal{L}}_1$  under $S$ implies the matrix equation
\bea
\label{matrix equation}
S^T \cdot Y&=& Y,
\eea
which can  be only satisfied for a vanishing $Y$. Had we restricted the symmetry to just two factors, for example $\langle S_1, S_2\rangle$, this term would have been allowed. However, in this case corresponding to the flavor S-symmetry being an $\left(Z_2\right)^2$-symmetry, one can check that the charged lepton squared mass matrix, proportional to $Y \cdot Y^\dagger$ when the SM Higgs gets its vacuum expectation value (vev), would be singular with two zero eigen values, and the charged lepton mass hierarchy can not thus be produced. Another serious drawback for the existence of such a term in $\left(Z_2\right)^2$-symmetry is that the left handed charged leptons need here to be rotated non-trivially in order to achieve the diagonalization of the charged lepton mass matrix, and this in turn would destroy the prediction of the NTB pattern in Eq.~(\ref{VNTB0}).

In order to remedy this, we add three SM-singlet scalar fields  $\Delta_k$ coupled to the lepton LH doublets through the
 dimension 5 operator:
\begin{eqnarray}
{\cal{L}}_2 &=& \frac{f_{ikr}}{\Lambda} \overline{L}_i \Phi \Delta_k  l^c_r \, .
\end{eqnarray}
and we assume $\Delta_k$ to vary under S-symmetry as \begin{equation}
 \Delta_i \rightarrow S_{i j}\Delta_{j}
\end{equation}
As in \cite{z2}, this {\it{ad hoc}} assumption of ${\cal{L}}_2$ with only one Higgs field is suitable to reduce the effects of
flavor changing neutral currents \cite{BjorWein77}. Invariance of ${\cal{L}}_2$ under S-symmetry leads to:
 \bea
S^T \, f_r \, S =f_r \eea
where $(f_r)_{ij}=f_{i j r}$. Thus, from Eq. \ref{equivalence}, the matrix $f_r$ has the NTB modified tripartite form:
\begin{eqnarray}
\label{cly} f_r=\pmatrix{A^r_0-B^r_0+C^r_0 & \frac{7\sqrt{2}}{47}(B^r_0-22C^r_0) & \frac{7\sqrt{2}}{47}(B^r_0+25C^r_0)  \cr \frac{7\sqrt{2}}{47}(B^r_0-22C^r_0) & A^r_0+C^r_0
&  \frac{1}{47}(49B^r_0+191C^r_0)\cr  \frac{7\sqrt{2}}{47}(B^r_0+25C^r_0) &  \frac{1}{47}(49B^r_0+191C^r_0) & A^r_0-C^r_0} .\end{eqnarray}
 When $\Delta_k$ and $\phi^\circ$ take the vevs
$<\Delta_k >=\delta_k$, $v=<\phi^\circ>$, then ${\cal{L}}_2$ would generate charged lepton mass
 matrix:
\begin{equation}
\left( {\cal{M}}_l \right)_{ir} = \frac{v f_{ikr}}{\Lambda}\delta_k
\end{equation}
One can arrange the vevs and the Yukawa couplings such that ${\cal{M}}_l$, after suitably rotating the flavor- and SM-singlets $l^c_j$,
is the charged lepton mass matrix in the flavor basis. For example, if   $\delta_1,\delta_2 \ll \delta_3$ we get
 \begin{eqnarray}\label{clm}
 M_l &=& \frac{v\delta_3}{\Lambda} \left ( \matrix{ A'_1  & A'_2 & A'_3 \cr B'_1 &
B'_2 & B'_3 \cr C'_1 & C'_2 & C'_3 \cr} \right ),
\end{eqnarray}
where
\bea
&A'_i = \frac{7\sqrt{2}}{47} (B^i_0+25C^i_0) \;\;\;\;\;\;,\;\;\;\;\;\; B'_i = \frac{1}{47} (49B^i_0+191C^i_0)\;\;\;\;\;\;,\;\;\;\;\;\;
C'_i = A^i_0-C^i_0 &
\eea
The charged lepton squared mass matrix $(M_l\;M_l^\dagger)$ assumes the form,
\bea
M_l\;M_l^\dagger
    &\approx&\left(\frac{v\delta_3}{\Lambda}\right)^2
    \pmatrix {{\bf A'} \cdot {\bf A'} & {\bf A'} \cdot {\bf B'} & {\bf A'}\cdot {\bf C'} \cr
     {\bf B'} \cdot {\bf A'} & {\bf B'} \cdot {\bf B'} & {\bf B'} \cdot  {\bf C'} \cr
     {\bf C'} \cdot {\bf A'} & {\bf C'} \cdot {\bf B'} & {\bf C'} \cdot {\bf C'}},
\eea
where ${\bf A'}$ is the complex vector of components $A'_i$ ($i=1,2,3$, similarly  for ${\bf B',C'}$) and the usual inner product
of two complex vectors (${\bf A'}$ and ${\bf B'}$) is defined as ${\bf A'} \cdot {\bf B'} \equiv \sum_{i=1}^{3} {A'}_i\, {B'}_i^* $.
The charged lepton mass matrix, $M_l$,  is nonsingular provided the three vectors  (${\bf A',B',C'}$) are linearly independent
(amounting to non-coplanar vectors in the real Euclidian case).

In order to show that $M_l$ can naturally represent the lepton mass matrix in the flavor basis, let us just assume
the magnitudes of the three vectors coming in ratios comparable to the lepton mass ratios:
\bea
\frac{|{\bf A'}|}{|{\bf C'}|} \equiv \la_e \sim \frac{m_e}{m_\tau} = 2.8 \times 10^{-4}&,&
\frac{|{\bf B'}|}{|{\bf C'}|} \equiv \la_\m \sim \frac{m_\m}{m_\tau} = 5.9 \times 10^{-2},
\eea
where $|{\bf A'}|$ represent the norm of the vector defined by,
\be |{\bf A'}|= \sqrt{{\bf A'} \cdot {\bf A'}}, \;\;\; (\mbox{similarly for}\;  {\bf B',C'}).
\ee
This leads the squared mass matrix
to be written as:
\bea
Q_\lam \equiv M_l M_l^\dagger &\approx& \left(\frac{v\delta_3}{\Lambda}\right)^2\; |{\bf C'}|^2
\left ( \matrix{
\lam_e^2 & \lam_e \lam_\mu \cos\psi\; e^{i\a} & \lam_e \cos \phi\; e^{i\b} \cr
\lam_e \lam_\mu \cos\psi\; e^{-i\a} & \lam_\mu^2 & \lam_\mu \cos \theta \;e^{i\g} \cr
\lam_e \cos \phi\; e^{-i\b} & \lam_\mu \cos \theta\; e^{-i\g} & 1 \cr} \right ),
\label{Q}
\eea
where $\psi$, $\phi$ and $\theta$ are the ``angles'' between the pairs of complex vectors ${(\bf A',B')},(\bf A',C')$ and $(\bf C',B')$ respectively,
whereas $\a$, $\b$ and $\g$ are the phases of the corresponding inner products \footnote{The ``angle'' $\psi$  and "phase" $\al$ between two complex vectors
${\bf A'}$ and ${\bf B'}$ are  defined, following Cauchy-Schwartz inequality, as $|{\bf A'} \cdot {\bf B'}|=|{\bf A'}|\;|{\bf B'}|\;\cos{\psi}$, so we have
${\bf A'}.{\bf B'}=|{\bf A'}|\; |{\bf B'}|\;\cos{\psi}\;e^{i\, \al}$, where $\al = \arg\left({\bf A'} \cdot {\bf B'}\right)$.}.
The diagonalization of $M_l M_l^\dagger$ by
means of an infinitesimal ``rotation'' amounts to seeking an antihermitian matrix
\be
I_\eps = \left ( \matrix{
0 & \eps_1 & \eps_2 \cr -\eps_1^* & 0 & \eps_3 \cr -\eps_2^* & -\eps_3^* & 0 \cr} \right ),
\ee
with small parameters $\eps'$s,
satisfying: \bea \left( Q_\lam + \left[Q_\lam,I_\eps \right] \right)_{ij}&=0,& i\neq j.
\eea
If we solve this equation analytically to express the $\eps$'s in terms of ($\lam_{e,\mu},\;\cos (\psi,\phi,\theta),\;\a,\;\b,\;\g $), we find, apart from
``fine tuned'' situations corresponding to coplanar vectors ${\bf A',B',C'}$, that we get:
$|\eps_3| \sim \lam_\mu,\; |\eps_2| \sim \lam_e$ and
$|\eps_1| \sim \lam_e/\lam_\mu$, which points to a consistent solution diagonalizing  $Q_\la$ close to the identity
matrix given by $U^l_L = e^{I_\eps} \approx I + I_\eps$. The eigenvalues for $M_l M_l^\dagger$ can be approximated up
to leading order in $\lam$'s, and by identifying them with the observed charged lepton squared masses  we get,
\bea
m_e^2 &=& {\mu^2\,\la_e^2\over \sin^2{\th}}\,\left(1-2\,\cos^2{\psi} +6\,\cos{\left(\d-\b-\al\right)}\,\cos{\psi}\,\cos{\th}\,\cos{\phi}-\cos^2{\th}
- 2\,\cos^2{\phi}\,\cos^2{\th} - 2\,\cos^2{\phi}\right),\nonumber\\
m_\mu^2&=& {\mu^2\,\la_\mu^2\over \sin^2{\th}}\,\left(1-3\,\cos^2{\th} + 2\cos^4{\th}\right)\nonumber\\
&& -{\mu^2\,\la_e^2\over \sin^2{\th}}\,\left(1-2\,\cos^2{\psi} +6\,\cos{\left(\d-\b-\al\right)}\,\cos{\psi}\,\cos{\th}\,\cos{\phi}-\cos^2{\phi}
- 2\,\cos^2{\phi}\,\cos^2{\th} - 2\,\cos^2{\th}\right),\nonumber\\
m_\tau^2 &=& \mu^2\,\left(1+ 2\,\la_\mu^2\,\cos^2{\th} + 2\,\la_e^2\,\cos^2{\phi}\right),
\eea
where $\mu =\left(\frac{v\delta_3}{\Lambda}\right)\; |{\bf C'}|$.
By giving some fixed values to the angles and phases, one can in general solve the above mentioned equations for $(\la_\mu, \la_e,\; \mbox{and}\; \mu)$.
 For illustrative purpose, we choose a common value $\pi/3$ for angles but for phases we have  representative values  as ($\a={\pi\over 3},\b={\pi\over 4}, \mbox{and}\; \g={\pi\over 5}$). The resulting solutions for $(\la_\mu, \la_e,\; \mbox{and}\; \mu)$ are,
 \bea
 \la_e= 7.99\,\times 10^{-4}, & \la_\mu = 8.4\,\times 10^{-2}, & \mu = 1776.978\; \mbox{MeV},
 \eea
while the `exact' unitary diagonalizing matrix is given by:
\be
U^l_L \sim
\left ( \matrix{
1 & 5.46 \times 10^{-3} \, \exp{\left(0.693\,i\,\pi\right)} &
5.41\times10^{-4}\,\exp{\left(0.126\,i\,\pi\right)} \cr
-5.46 \times 10^{-3} \, \exp{\left(-0.693\,i\,\pi\right)} & 1 &
 4.23 \times 10^{-2}\,\exp{\left( 0.20\,i\,\pi\right)} \cr
 -5.41\times10^{-4}\,\exp{\left(-0.126\,i\,\pi\right)}  &
 -4.23 \times 10^{-2}\,\exp{\left(- 0.20\,i\,\pi\right)} & 1 \cr} \right ).
 \label{infunit}
 \ee
The deviations due to the rotations are generally small, and thus produce  tiny but acceptable modifications to the mixing and phase angles
of the unitary mixing matrix in Eq.~(\ref{VNTB0}).

\section{The NTB neutrino mass matrix and type-I seesaw scenario }
Here, the type-I seesaw formula (Eq. \ref{Majmass}) applies. The Dirac neutrino mass matrix comes from
\begin{eqnarray} g_{ij} \overline{L}_i
\tilde{\Phi} \nu_{Rj},
\end{eqnarray}
where $\tilde{\Phi} = i \tau_2 \Phi^*$, and we assume the RH neutrinos transforming under $S$ as
\begin{eqnarray} \label{rtransform}
\nu_{Rj} \rightarrow S_{j\gamma} \nu_{R\gamma},
\end{eqnarray}

Then, the invariance of the Lagrangian under S-symmetry leads to
\begin{eqnarray}
S^T. g.S=g .
\end{eqnarray}
Thus we have the NTB modified tripartite form for $g$, and when $\tilde{\Phi}$ takes a vev we obtain the Dirac mass matrix as:
\begin{eqnarray}
\label{nDm} M_{\nu 0}^D &=& v\pmatrix{A^D_0-B^D_0+C^D_0 & \frac{7\sqrt{2}}{47}(B^D_0-22C^D_0) & \frac{7\sqrt{2}}{47}(B^D_0+25C^D_0)  \cr \frac{7\sqrt{2}}{47}(B^D_0-22C^D_0) & A^D_0+C^D_0
&  \frac{1}{47}(49B^D_0+191C^D_0)\cr  \frac{7\sqrt{2}}{47}(B^D_0+25C^D_0) &  \frac{1}{47}(49B^D_0+191C^D_0) & A^D_0-C^D_0} .\end{eqnarray}

Again, the invariance under S-symmetry of the term $\frac{1}{2} \nu^T_{iR} C \left( M_R \right)_{ij} \nu_{jR}$ ($C$ is the charge conjugation matrix)
would impose the NTB modified tripartite form for the Majorana RH neutrino mass matrix:
\begin{eqnarray}
\label{nRm}
M_{R0} &=& \Lambda_R \pmatrix{A^R_0-B^R_0+C^R_0 & \frac{7\sqrt{2}}{47}(B^R_0-22C^R_0) & \frac{7\sqrt{2}}{47}(B^R_0+25C^R_0)  \cr \frac{7\sqrt{2}}{47}(B^R_0-22C^R_0) & A^R_0+C^R_0
&  \frac{1}{47}(49B^R_0+191C^R_0)\cr  \frac{7\sqrt{2}}{47}(B^R_0+25C^R_0) &  \frac{1}{47}(49B^R_0+191C^R_0) & A^R_0-C^R_0}.
 \end{eqnarray}
where $\Lambda_R$ is a high scale characterizing the seesaw mechanism.

Thus we get also an NTB modified tripartite form for the effective neutrino mass matrix:
\begin{eqnarray}\label{nm} M_{\nu 0} &=&
\frac{v^2}{\Lambda_R} \pmatrix{A_{\n 0}-B_{\n 0}+C_{\n 0} & \frac{7\sqrt{2}}{47}(B_{\n 0}-22C_{\n 0}) & \frac{7\sqrt{2}}{47}(B_{\n 0}+25C_{\n 0})  \cr \frac{7\sqrt{2}}{47}(B_{\n 0}-22C_{\n 0}) & A_{\n 0}+C_{\n 0}
&  \frac{1}{47}(49B_{\n 0}+191C_{\n 0})\cr  \frac{7\sqrt{2}}{47}(B_{\n 0}+25C_{\n 0}) &  \frac{1}{47}(49B_{\n 0}+191C_{\n 0}) & A_{\n 0}-C_{\n 0}},
 \end{eqnarray}
where $A_{\n 0}, B_{\n 0}$ and $C_{\n 0}$ are given by,
\bea
A_{\n 0} &=& {13\over75} \frac{\left( A^D_0-\frac{49}{47}B^D_0+\frac{279}{47}C^D_0\right)^2}{A^R_0-\frac{49}{47}B^R_0+\frac{279}{47}C^R_0} + {101\over 300} \frac{\left( A^D_0-\frac{49}{47}B^D_0-\frac{426}{47}C^D_0\right)^2}{A^R_0-\frac{49}{47}B^R_0-\frac{426}{47}C^R_0} + {49\over 100}\frac{\left( A^D_0+\frac{51}{47}B^D_0+\frac{194}{47}C^D_0\right)^2}{A^R_0+\frac{51}{47}B^R_0+\frac{194}{47}C^R_0},
\nonumber\\\nonumber\\
B_{\n 0} &=& -{31\over75} \frac{\left( A^D_0-\frac{49}{47}B^D_0+\frac{279}{47}C^D_0\right)^2}{A^R_0-\frac{49}{47}B^R_0+\frac{279}{47}C^R_0} - {17\over 300} \frac{\left( A^D_0-\frac{49}{47}B^D_0-\frac{426}{47}C^D_0\right)^2}{A^R_0-\frac{49}{47}B^R_0-\frac{426}{47}C^R_0} + {47\over 100}\frac{\left( A^D_0+\frac{51}{47}B^D_0+\frac{194}{47}C^D_0\right)^2}{A^R_0+\frac{51}{47}B^R_0+\frac{194}{47}C^R_0},
\nonumber\\\nonumber\\
C_{\n 0} &=& {1\over 15} \frac{\left( A^D_0-\frac{49}{47}B^D_0+\frac{279}{47}C^D_0\right)^2}{A^R_0-\frac{49}{47}B^R_0+\frac{279}{47}C^R_0} - {1\over 15} \frac{\left( A^D_0-\frac{49}{47}B^D_0-\frac{426}{47}C^D_0\right)^2}{A^R_0-\frac{49}{47}B^R_0-\frac{426}{47}C^R_0}.
\label{relabc}
\eea

The eigenvalues of the light neutrino mass matrix in Eq.~(\ref{nm}) are found to be:
\bea
\label{mspec}
m_1 = \frac{v^2}{\Lambda_R} \frac{\left( A^D_0-\frac{49}{47}B^D_0+\frac{279}{47}C^D_0\right)^2}{A^R_0-\frac{49}{47}B^R_0+\frac{279}{47}C^R_0},
& m_2 = \frac{v^2}{\Lambda_R} \frac{\left( A^D_0-\frac{49}{47}B^D_0-\frac{426}{47}C^D_0\right)^2}{A^R_0-\frac{49}{47}B^R_0-\frac{426}{47}C^R_0},&
m_3 = \frac{v^2}{\Lambda_R} \frac{\left( A^D_0+\frac{51}{47}B^D_0+\frac{194}{47}C^D_0\right)^2}{A^R_0+\frac{51}{47}B^R_0+\frac{194}{47}C^R_0}.\nonumber\\  \eea
As mentioned before (see Eq.~\ref{eigen-trip-0}), but now specifying to the seesaw model, all types of neutrino mass hierarchies can be accommodated  as follows.
\begin{itemize}
\item{ Normal hierarchy}: with
\begin{eqnarray} A^i_0 \simeq
B^i_0,\; C^i_0 \ll B^i_0,\; i=R,D
\end{eqnarray}
we get a mass spectrum corresponding to normal hierarchy:
\begin{eqnarray}
[m_1,m_2,m_3] &\simeq& \frac{v^2}{\Lambda_R} \left[ \frac{(C^D_0)^2}{C^R_0},\; \frac{(C^D_0)^2}{C^R_0},\; \frac{(A^D_0)^2}{A^R_0} \right]
\end{eqnarray}
\item{ Inverted hierarchy}: with
\begin{eqnarray} A^i_0 \simeq
- B^i_0,\; C^i_0 \ll B^i_0,\; i=R,D
\end{eqnarray}
we get a mass spectrum corresponding to inverted hierarchy:
\begin{eqnarray}
[m_1,m_2,m_3] &\simeq& \frac{v^2}{\Lambda_R} \left[ \frac{(A^D_0)^2}{A^R_0},\; \frac{(A^D_0)^2}{A^R_0},\; \frac{(C^D_0)^2}{C^R_0} \right]
\end{eqnarray}
\item{ Degenerate case}: with
\begin{eqnarray}
A_i \gg B_i \gg C_i,\;
i=R,D\end{eqnarray}
we get a quasi degenerate mass spectrum
\begin{eqnarray}
[m_1,m_2,m_3] &\simeq& \frac{v^2}{\Lambda_R} \left[ \frac{(A^D_0)^2}{A^R_0},\; \frac{(A^D_0)^2}{A^R_0},\; \frac{(A^D_0)^2}{A^R_0} \right]
\end{eqnarray}
\end{itemize}
\vspace{-3mm}

The RH neutrino mass term violates lepton number by two units, and could be a source of lepton asymmetry. The produced asymmetry due to the out of equilibrium decay of the lightest RH  neutrino to SM particles is given by \cite{FY}:
\begin{eqnarray}
\epsilon \simeq \frac{3}{16\pi
v^2}\frac{1}{\left(\tilde{M}_\nu^{D\dagger}\tilde{M}_\nu^D\right)_{11}}\sum_{j=2,3}
\mbox{Im}\left\{\left[(\tilde{M}_\nu^{D\dagger}\tilde{M}_\nu^D)_{1j}\right]^2\right\}\frac{M_{R1}}{M_{Rj}},
\end{eqnarray}
where $M_{Ri},\; i=1\cdots 3$ are the masses for RH
neutrinos, and $\tilde{M}_\nu^D$ is the Dirac neutrino mass
matrix in the basis where the Majorana RH neutrino mass
matrix $M_{R 0}$ is diagonal.
Since the RH neutrino mass matrix (Eq. \ref{nRm}) has the NTB modified tripartite form, then it is diagonalized by $V^{NTB}_0$ (Eq. \ref{equivalence}). Thus under $\n_R \rightarrow V^{NTB}_0\; \n_R$ we have $M^D_{\n 0} \rightarrow M^D_{\n 0}\; V^{NTB}_0$. We still have
freedom in multiplying the diagonalizing unitary matrix $V^{NTB}_0$ by diagonal phases $F_0 = \mbox{diag}\left(e^{-i \a_1},\; e^{-i \a_2},\;e^{-i \a_3} \right)$ adjusted  normally so that
to cancel the phases of the spectrum of $M_{R 0}$. Namely, these phases cancel out upon choosing
\bea
\left( \a_1, \a_2, \a_3 \right) &=& \frac{1}{2}\mbox{Arg} \left( A^R_0-\frac{49}{47}B^R_0+\frac{279}{47}C^R_0,\;
A^R_0-\frac{49}{47}B^R_0-\frac{426}{47}C^R_0,\; A^R_0+\frac{51}{47}B^R_0+\frac{194}{47}C^R_0
\right) \eea
Thus, we have $\tilde{M}_{\nu 0}^D = M _{\nu 0}^D \cdot V^{NTB}_0 \cdot F_0$ so that we can write
\bea
\tilde{M}_{\nu 0}^{D\dagger}\;\tilde{M}_{\nu 0}^D &=& F_0^\dagger\; V^{NTB \dagger}_0\; M_{\n 0}^{D \dagger}\; V^{NTB}_0\;
V^{NTB ^\dagger}_0\;   M_{\n 0}^D\; V^{NTB}_0\; F_0
\eea
which shows that $\tilde{M}_{\nu 0}^{D\dagger}\tilde{M}_{\nu 0}^D$ is
diagonal and real. Whence $\epsilon$ vanishes in this scenario if S-symmetry is exact.

\section{The NTB neutrino mass matrix and type-II seesaw scenario}
We apply in this section the type-II seesaw scenario aiming to show that it can solely accommodate an enough lepto/baryogenesis
for the observed baryon/photon density in the universe. 
As in \cite{z2}, we introduce in this scenario two SM triplet fields $\Sigma_A$, $A=1,2$ which are singlet
under the flavor S-symmetry. The Lagrangian part relevant for the neutrino mass matrix is:
\begin{equation}
{\cal{L} } = \lambda_{\alpha\beta}^{A}\, L_\alpha^T\, C\, \Sigma_A\, i\,\tau_2\, L_\beta +
{\cal{L}}(H,\Sigma_A)+h.c.
\end{equation}
where $A=1,2$ and
\begin{eqnarray}
{\cal{L}}(H,\Sigma_A) &=& \mu_H^2 H^\dagger H + \frac{\lambda_H}{2} {(H^\dagger H)}^2+
M_A\, \mbox{Tr}\left(\Sigma_A^\dagger \Sigma_A \right)+
\frac{\lambda_{\Sigma_A}}{2} \left[\mbox{Tr}\left( \Sigma_A^\dagger \Sigma_A\right)\right]^2
+ \\\nonumber && \lambda_{H\Sigma_A} (H^\dagger H) \mbox{Tr}\left( \Sigma^\dagger_A \Sigma_A\right)
+
{\mu_A H^T \Sigma_A^\dagger i\tau_2 H +h.c.}
\end{eqnarray}
where $H$ and $\Sigma_A$ are written as
\bea
H=\pmatrix{\phi^+ \cr \phi^0}, &&  \Sigma_A = \left ( \matrix{ \frac{\Sigma^+}{\sqrt{2}}  &
\Sigma^0 \cr \Sigma^{++} & -\frac{\Sigma^+}{\sqrt{2}}\cr} \right )_A.
\eea

The neutrino mass matrix due to the exchange of the two triplets, $\Sigma_1$ and $\Sigma_2$, is
\begin{equation} \label{mass}
(M_\nu)_{\alpha\beta}\simeq v^2 \left[\lambda^1_{\alpha\beta} \frac{\mu_1}{M^2_{\Sigma_1}} +
\lambda^2_{\alpha\beta}\frac{\mu_2}{M^2_{\Sigma_2}}\right]
\end{equation}
where $M_{\Sigma_i}$ is the mass of the neutral component $\Sigma_i^0$ of the triplet $\Sigma_i ,i=1,2$.

Appropriately, we present some remarks here. First,  the flavor S-symmetry would force the matrices $\lambda^1$ and $\lambda^2$ to have the same NTB modified tripartite structure:
 \bea
 \lam^A_0 = \pmatrix{A^A_0-B^A_0+C^A_0 & \frac{7\sqrt{2}}{47}(B^A_0-22C^A_0) & \frac{7\sqrt{2}}{47}(B^A_0+25C^A_0)  \cr \frac{7\sqrt{2}}{47}(B^A_0-22C^A_0) & A^A_0+C^A_0
&  \frac{1}{47}(49B^A_0+191C^A_0)\cr  \frac{7\sqrt{2}}{47}(B^A_0+25C^A_0) &  \frac{1}{47}(49B^A_0+191C^A_0) & A^A_0-C^A_0},\;A=1,2
\eea
Thus, the neutrino mass matrix in Eq. (\ref{mass}) has the NTB modified tripartite form, hence it can generate all types of
neutrino mass hierarchies (Eqs. \ref{nh}, \ref{ih}, \ref{dh}).

 Second, the $\mu_A$-term in ${\cal{L}}(H,\Sigma_A)$, which does not allow an `undesirable' spontaneous breaking of the lepton number, permits to arrange the parameters so that minimizing the potential gives
a non-zero vev for the neutral component $\Sigma^0$ of the triplet. This would generate the mass term for the neutrinos in Eq. (\ref{mass}) in an equivalent way to integrating out the heavy triplets. Third, the flavor changing  neutral current due to the triplet is highly
suppressed because of the large value of its
mass scale (or equivalently due to  the smallness of the neutrino masses).

One can discuss now the baryon asymmetry generated by leptogenesis. We show at present that even though the neutrino Yukawa
couplings are real
it is possible to generate a  baryon to photon density consistent with the observations. In fact, since the triplet
$\Sigma_A$ can decay into lepton pairs $L_\alpha L_\beta$ and $HH$, it implies that these processes violate
total lepton numbers (by two units) and may establish a lepton asymmetry. As the universe cools further,
the sphaleron interaction \cite{KRS} converts this asymmetry into baryon asymmetry. At temperature of the
order $\mbox{max}\{M_1, M_2\}$, the heaviest triplet would decay via  lepton number  violating interactions.
Nonetheless, no asymmetry will be generated from this decay since the rapid  lepton number violating interactions
due to the lightest Higgs triplet will erase any previously generated  lepton asymmetry. Therefore, only when the
temperature becomes just below the mass of the lightest triplet Higgs the asymmetry would be generated.

With just one triplet, the lepton
asymmetry will be generated at the two loop level and it is highly suppressed. We justify this in that one can
always redefine the phase of the Higgs field to make the $\mu$ real resulting in the absorptive part of the self energy diagram
becoming equal to zero. The choice of having more than one Higgs triplet is
necessary to generate the asymmetry \cite{HMS}. In this case,  the CP asymmetry in the decay of the lightest
Higgs triplet (which we choose to be $\Sigma_1$) is generated at one loop level due to the interference
between the tree and the one loop self energy diagram, and is given by
\begin{equation}
\epsilon_{CP} \approx -\frac{1}{8\pi^2} \frac{\mbox{Im}\left[\mu_1\mu_2^*
\mbox{Tr}\left(\lambda^1\lambda^{2\dagger}\right)\right]}{M_2^2}
\frac{M_1}{\Gamma_1},
\end{equation}
where $\Gamma_1$ is the decay rate of the lightest Higgs triplet and it is given by
\begin{eqnarray}
\Gamma_1 = \frac{M_1}{8\pi}\left[\mbox{Tr}\left(\lambda^{1\dagger} \lambda^1\right)
+ \frac{\mu_1^2}{M_1^2}   \right].
\end{eqnarray}

The baryon to photon density is approximately given by
\begin{equation}
\eta_B \equiv \frac{n_B}{s} =\frac{1}{3}\eta_L \simeq \frac{1}{3} \frac{1}{g_* K} \epsilon_{CP},
\end{equation}
where $g_* \sim 100$ is the number of relativistic degrees of freedom at the time when the Higgs
triplet decouples from the thermal bath and $K$ is the efficiency factor  which takes into account the fraction of out-of equilibrium decays and the washout effect. In the case of strong wash out, the efficiency factor can be approximated by ($H$ is the Hubble parameter)
\begin{eqnarray}
K \simeq \frac{\Gamma_1}{H}(T= M_1),
\end{eqnarray}
 For $|\mu_{1,2}| \approx M_{\Sigma_{1,2}} \sim 10^{12}\, \mbox{GeV}$, and an efficiency factor of order $K \sim 10^{4}$, and assuming
 real matrices $\lambda^A$, and denoting the phases of $\m_A$ by $\phi_A$ we find

\begin{equation}
\eta_B\approx  10^{-7}\frac{
\mbox{Tr}\left(\lambda^1\lambda^{2\dagger}\right)}{\mbox{Tr}\left(\lambda^{1\dagger} \lambda^1 \right) + 1}
\sin(\phi_2 -\phi_1)
\end{equation}
Thus one can produce the correct baryon-to photon ratio of $\eta_B \simeq 10^{-10}$ by choosing $\lambda$'s of order $0.1$ and not too small relative phase between the $\mu$'s.

\section{Summary and conclusion}
We derived an explicit realization of the $\left(Z_2\right)^3$ symmetry characterizing uniquely the non-tri-bimaximal pattern with $\t_z \neq 0$, in line with the recent neutrino oscillations data.
This would constitute a natural explanation for the observed neutrino mixing, rather than considering the tri-bimaximal as the
zero approximation followed by some kind of perturbation in order to fit the observed mixing. Actually, the recent oscillation data make the perturbative treatment implausible since large deviations are needed in order  to fit the mixing angle $\t_z$. We have imposed the $\left(Z_2\right)^3$ symmetry in a setup including charged leptons, neutrinos and
extra scalar fields in order to account for the charged lepton mass hierarchy. In type-I seesaw scenario, we could account for the various neutrino mass hierarchies. Finally, in type-II seesaw scenario, we could by choosing appropriate Yukawa couplings interpret the observed baryon to photon ratio observed in the universe.

\section*{Acknowledgements}
N.C. acknowledges funding provided by the Alexander von Humboldt Foundation. Part of the work was done
within the associate scheme program of ICTP.



\bibliographystyle{mdpi}

\end{document}